\pdfoutput=1
\RequirePackage{ifpdf}
\ifpdf 
\documentclass[pdftex]{sigma}
\else
\documentclass{sigma}
\fi

\numberwithin{equation}{section}

\newtheorem{Theorem}{Theorem}[section]
\newtheorem{Lemma}[Theorem]{Lemma}
\newtheorem{Proposition}[Theorem]{Proposition}
\newtheorem{Conjecture}[Theorem]{Conjecture}
 { \theoremstyle{definition}
\newtheorem{Definition}[Theorem]{Definition}
\newtheorem{Remark}[Theorem]{Remark} }

\newcommand{\ii}{{\rm i}}

\newcommand{\dd}[1]{{\rm d}#1 }

\newcommand{\p}{\partial}

\begin{document}
\allowdisplaybreaks

\newcommand{\arXivNumber}{1810.07919}

\renewcommand{\PaperNumber}{001}

\FirstPageHeading

\ShortArticleName{Aspects of Hecke Symmetry: Anomalies, Curves, and Chazy Equations}

\ArticleName{Aspects of Hecke Symmetry:\\
Anomalies, Curves, and Chazy Equations}

\Author{Sujay K.~ASHOK~$^\dag$, Dileep P.~JATKAR~$^\ddag$ and Madhusudhan RAMAN~$^\S$}

\AuthorNameForHeading{S.K.~Ashok, D.P.~Jatkar, and M.~Raman}

\Address{$^\dag$~Institute of Mathematical Sciences, Homi Bhabha National Institute (HBNI),\\
\hphantom{$^\dag$}~IV Cross Road, C.~I.~T.~Campus, Taramani, Chennai 600 113, India}
\EmailD{\href{mailto:sashok@imsc.res.in}{sashok@imsc.res.in}}

\Address{$^\ddag$~Harish-Chandra Research Institute, Homi Bhabha National Institute (HBNI),\\
\hphantom{$^\ddag$}~Chhatnag Road, Jhunsi, Allahabad 211 019, India}
\EmailD{\href{mailto:dileep@hri.res.in}{dileep@hri.res.in}}

\Address{$^\S$~Department of Theoretical Physics, Tata Institute of Fundamental Research,\\
\hphantom{$^\S$}~Homi Bhabha Road, Navy Nagar, Colaba, Mumbai 400 005, India}
\EmailD{\href{mailto:madhur@theory.tifr.res.in}{madhur@theory.tifr.res.in}}

\ArticleDates{Received May 06, 2019, in final form December 29, 2019; Published online January 01, 2020}

\Abstract{We study various relations governing quasi-automorphic forms associated to discrete subgroups of ${\rm SL}(2,\mathbb{R}) $ called Hecke groups. We show that the Eisenstein series associated to a Hecke group ${\rm H}(m)$ satisfy a set of $m$ coupled linear differential equations, which are natural analogues of the well-known Ramanujan identities for quasi-modular forms of~${\rm SL}(2,\mathbb{Z})$. Each Hecke group is then associated to a (hyper-)elliptic curve, whose coefficients are determined by an anomaly equation. For the $m=3$ and $4$ cases, the Ramanujan identities admit a natural geometric interpretation as a Gauss--Manin connection on the parameter space of the elliptic curve. The Ramanujan identities also allow us to associate a nonlinear differential equation of order~$ m $ to each Hecke group. These equations are higher-order analogues of the Chazy equation, and we show that they are solved by the quasi-automorphic Eisenstein series $E_2^{(m)}$ associated to ${\rm H}(m) $ and its orbit under the Hecke group. We conclude by demonstrating that these nonlinear equations possess the Painlev\'e property.}


\Classification{34M55; 11F12; 33E30}

\vspace{-1mm}

\section{Introduction}

In this work, we study properties of automorphic forms for Hecke groups. An element of the Hecke group, which we denote ${\rm H}(m)$, is any word made up of the letters $ S $ and $ T $ that is further circumscribed by the relations
\begin{gather*}
S^2 = 1, \qquad (ST)^m = 1.
\end{gather*}
It is easy to see that for $ m = 3 $, the generators of the Hecke group satisfy the same relations that bind the generators of the modular group. More concretely, for $ \tau $ that takes values in the upper-half plane $\mathbb{H} $, the generators of the Hecke group act as\footnote{Hecke groups are usually defined in the literature as generated by the symbols $ T $ and $ S $ that act as
\begin{gather*}
T \colon \ \tau \rightarrow \tau + \sqrt{\lambda_m} \qquad \text{and} \qquad S \colon \ \tau \rightarrow -\frac{1}{\tau},
\end{gather*}
where $ \lambda_m $ is defined as in \eqref{eq:heckeSpectrum}. It is easily verified that under the rescaling $ \tau \rightarrow \sqrt{\lambda_m} \tau $, we recover the definition provided in the main text.}
\begin{gather*}
T \colon \ \tau \rightarrow \tau + 1, \qquad S \colon \ \tau \rightarrow -\frac{1}{\lambda_m \tau},
\end{gather*}
where
\begin{gather}\label{eq:heckeSpectrum}
\lambda_m = 4\cos^2\left(\frac{\pi}{m} \right).
\end{gather}
The Hecke groups are indexed by an integer $ m \geq 3 $ that will be called its \emph{height}. For $ m \in \{ 3,4,6,\infty\} $ the corresponding $ \lambda_m $ are integers. These groups are called \emph{arithmetic} Hecke groups, while the rest (for which $ \lambda_m $ is not an integer) will be referred to as \emph{non-arithmetic} Hecke groups~\cite{Shen2016}. In the interest of uniformity, we will restrict our attention to Hecke groups with finite heights.

In Section~\ref{sec:Ramanujan}, we begin with a brief review of \cite{Doran:2013npa} on (quasi-)automorphic forms for Hecke groups~${\rm H}(m)$. The ring of quasi-automorphic forms for the Hecke group~${\rm H}(m)$, which we denote~$ {\mathbf {\mathcal R}_m} $, is generated by the Eisenstein series $E_{2k}^{(m)}$ associated to the Hecke group, which in turn are related to solutions of a generalized Halphen system in a simple way. The Eisenstein series are found to satisfy a system of~$m$ first-order coupled linear differential equations which are natural analogues of the Ramanujan identities corresponding to the modular group ${\rm H}(3)$
\begin{gather*}
\frac{1}{2\pi\ii} \frac{\dd}{\dd\tau}E_2^{(3)} = \frac{1}{12} \big(\big(E_2^{(3)}\big)^2 - E_4^{(3)}\big), \\
\frac{1}{2\pi\ii} \frac{\dd}{\dd\tau}E_4^{(3)} = \frac{1}{3} \big(E_2^{(3)}E_4^{(3)} - E_6^{(3)}\big), \\
\frac{1}{2\pi\ii} \frac{\dd}{\dd\tau}E_6^{(3)} = \frac{1}{2} \big(E_2^{(3)}E_6^{(3)} - \big(E_4^{(3)}\big)^2\big).
\end{gather*}

In Section \ref{sec:CurvesAnomalies}, we associate an algebraic curve to the Hecke group ${\rm H}(m)$, whose coefficients~$ A_k^{(m)} $ are quasi-automorphic forms.\footnote{As we will see, for $ m =3, 4 $, this curve is elliptic, and for $ m > 4 $ it is hyperelliptic.} An anomaly equation governing these coefficients is derived; it plays a role analogous to modular anomaly equations that have appeared in the literature on supersymmetric gauge theories \cite{Billo:2015pjb,Billo:2015jyt,Huang:2013eja} in that it fixes the dependence of these coefficients on the quasi-automorphic Eisenstein series $ E_2^{(m)} $. A complete solution to these anomaly equations requires the specification of boundary conditions that fix the dependence of $ A_k^{(m)} $ on purely automorphic pieces; this is done by insisting on certain fall-offs near the cusp $\tau=\ii\infty$ which we term `cuspidal' boundary conditions.

The discussion of curves is developed with the goal of interpreting the Ramanujan identities as corresponding to some natural geometric object on the parameter space of these curves. This has already been done for the modular group in \cite{Dubrovin:1994,Katz:1976,Movasati2012a,Movasati2012b}. We review and extend these results to the case of ${\rm H}(4)$ and construct a Ramanujan vector field that acts naturally on the period integrals of the associated curve. We also show that the discriminant of the curve is closely related to the automorphic discriminant defined in~\cite{Doran:2013npa}.

Another motivation for our studies comes from \cite{Takhtajan:1992qb}, where it was shown that for the modular group, the quasi-modular form $E_2^{(3)}$ and its ${\rm SL}(2,\mathbb{Z})$-orbits satisfy the well-known Chazy equation and further, that the weight $12$ modular discriminant plays the role of a Hirota $ \tau $-function for the Chazy equation. We use the integrable systems nomenclature here and by Hirota $ \tau $-function, we mean a solution to (possibly a collection of) nonlinear PDEs \cite{Jimbo1981b,Jimbo1981a,Takhtajan:1992qb}. In Section~\ref{sec:Chazy}, we observe that for every Hecke group the corresponding Eisenstein series $E_2^{(m)}$ satis\-fies an ordinary differential equation of order $m$ that can be systematically constructed using the Ramanujan identities. We go on to show that the ${\rm H}(m)$-orbit of $E_2$ also satisfies the same differential equation, and an analogue of the modular discriminant plays the role of a Hirota $ \tau $-function for the Hecke group, thereby generalizing the results of \cite{Takhtajan:1992qb} to all Hecke groups.

One of the important differences between the modular group and other Hecke groups is that there are non-trivial relations among the Eisenstein series of ${\rm H}(m)$ for $ m>3 $. These relations along with the Ramanujan identities naturally lead us to a third-order nonlinear differential equation, allowing us to make contact with earlier work on the subject by Maier~\cite{Maier}.

Finally, in Section \ref{sec:Painleve}, we perform a generalized Painlev\'e analysis of the higher-order Chazy equations and Maier's equation and show that all these equations possess the Painlev\'e property. Some technical material is collected in Appendices~\ref{app:Fourier} and~\ref{earlierwork}.

\section{Ramanujan identities}\label{sec:Ramanujan}
The automorphic forms we study in this paper are Eisenstein series corresponding to the Hecke group, denoted ${\rm H}(m)$. They will be built out of solutions to the generalized Halphen system, following~\cite{Doran:2013npa}. Once constructed, we prove analogues of the Ramanujan identities~-- first discussed in \cite{Ramanujan1916}~-- for all Hecke groups.

\subsection{Generalized Halphen systems and a proof}
 The generalized Halphen system is a set of coupled ordinary differential equations for three variables $ \lbrace t_k^{(m)}(\tau) \rbrace_{k=1}^{3} $ that satisfy
\begin{gather}
t_1^{{(m)}^\prime} =(a-1)\big(t_1^{(m)}t_2^{(m)}+t_1^{(m)}t_3^{(m)}-t_2^{(m)}t_3^{(m)}\big)+(b+c-1) \big(t_1^{(m)}\big)^2, \nonumber\\
t_2^{{(m)}^\prime} =(b-1)\big(t_2^{(m)}t_3^{(m)}+t_2^{(m)}t_1^{(m)}-t_1^{(m)}t_3^{(m)}\big)+(a+c-1) \big(t_2^{(m)}\big)^2, \nonumber\\
t_3^{{(m)}^\prime} =(c-1)\big(t_3^{(m)}t_1^{(m)}+t_3^{(m)}t_2^{(m)}-t_1^{(m)}t_2^{(m)}\big)+(a+b-1) \big(t_3^{(m)}\big)^2,\label{eq:GenHalphen}
\end{gather}
where the parameters $(a,b,c)$ are specified by the height $m$ of the Hecke group ${\rm H}(m)$ as
\begin{gather}
\label{eq:abcHalphen}
a =\frac{1}{2}\left(\frac{1}{2}+\frac{1}{m} \right),\qquad
b =\frac{1}{2}\left(\frac{1}{2}-\frac{1}{m}\right) ,\qquad
c = \frac{1}{2}\left(\frac{3}{2}-\frac{1}{m} \right) ,
\end{gather}
and the accent $ ' $ denotes the following derivative
\begin{gather} \label{eq:AHSII_Revised_V3:1}
f' := \frac{1}{2\pi\ii} \frac{\dd{}}{\dd{\tau}} f(\tau).
\end{gather}
It is at times more convenient to work with Fourier expansions of the automorphic forms we will introduce. The `nome' in our conventions is $ q = \text{e}^{2\pi\ii \tau} $, and when working with $ q $-series, the accent~$ ' $ is
\begin{gather*}
\frac{1}{2\pi\ii} \frac{\dd{}}{\dd{\tau}} = q \frac{\dd{}}{\dd{q}}.
\end{gather*}

The solution to the generalized Halphen system can be obtained explicitly in terms of hypergeometric functions whose arguments depend on the standard hauptmodul of the Hecke group \cite[Theorem~3(i)]{Doran:2013npa}. We have included a few details about these solutions and their Fourier expansions in Appendix~\ref{app:Fourier}.\footnote{It will be important to keep in mind that some of our normalizations differ from those of \cite{Doran:2013npa}.}

The Eisenstein series $ \big\lbrace E_{2k}^{(m)} \big\rbrace_{k=2}^{m} $ are holomorphic automorphic forms of weight $2k$ under the Hecke group ${\rm H}(m)$, and they have simple expressions in terms of the solutions to the generalized Halphen system \cite[see p.~707 and Theorem~4(iv)]{Doran:2013npa}. In order to simplify expressions, we define the linear combinations
\begin{gather}\label{eq:xydef}
\mathsf{x}^{(m)}= t_1^{(m)}-t_2^{(m)}\qquad \text{and}\qquad \mathsf{y}^{(m)} = t_3^{(m)}-t_2^{(m)}.
\end{gather}
The automorphic forms $E_{2k}^{(m)}$ in these variables are
\begin{gather}\label{eq:E2kxy}
E_{2k}^{(m)} = \big(\mathsf{x}^{(m)}\big)^{k-1} \mathsf{y}^{(m)}.
\end{gather}
It is clear from this resolution of the Eisenstein series into combinations of generalised Halphen variables that for $ m > 3 $, the algebra of automorphic forms is not freely generated by the $ \big\lbrace E_{2k}^{(m)} \big\rbrace_{k=2}^{m} $, as
\begin{gather}\label{eq:notFreelyGenerated}
E_{2p}^{(m)}E_{2(k-p+1)}^{(m)} = \big(\mathsf{x}^{(m)}\big)^{k-1}\big(\mathsf{y}^{(m)}\big)^2 = E_{2p'}^{(m)}E_{2(k-p'+1)}^{(m)}
\end{gather}
for \emph{any} integers $ 2 \leq p \leq k-1 $ and $ 2 \leq p' \leq k-1 $.

\looseness=-1 The Hecke groups also come equipped with a quasi-automorphic weight $ 2 $ Eisenstein series~$ E_2^{(m)} $, which is a linear combination of solutions to the generalized Halphen system \cite[Theorem~4(iii)]{Doran:2013npa}. In terms of the variables $ \big(\mathsf{x}^{(m)},\mathsf{y}^{(m)}\big) $ the Eisenstein series $E_2^{(m)}$ can be written as
\begin{gather}\label{eq:E2xy}
E_2^{(m)} = -\frac{1}{m-2} \big[ 4 \mathsf{x}^{(m)} + 2m \mathsf{y}^{(m)} + (3m+2) t_2^{(m)} \big],
\end{gather}
and using the conventions made explicit in Appendix \ref{app:Fourier}, we can check that as $ \tau \rightarrow \ii\infty $,
\begin{gather*}
E_2^{(m)} = 1 + O(q).
\end{gather*}

\begin{Lemma}The Ramanujan identities for Eisenstein series $ E_{2k}^{(m)} $ corresponding to Hecke groups ${\rm H}(m)$ take the form
\begin{gather}
E_{2}^{{(m)}^\prime} = \frac{m-2}{4m} \big( \big(E_2^{(m)}\big)^2 - E_4^{(m)} \big), \nonumber\\
E_{2k}^{{(m)}^\prime} = \frac{k}{2} \left( \frac{m-2}{m} \right)E_2^{(m)} E_{2k}^{(m)} -\left(\frac{k-2}{2}\right)E_{2k'}^{(m)} E_{2(k-k'+1)}^{(m)} - \left(\frac{m-k}{m}\right)E_{2k+2}^{(m)} ,\label{eq:RamanujanIdGeneral}
\end{gather}
for any $ k'$ such that $ 2 \leq k' \leq k-1 $.
\end{Lemma}

\begin{proof}Perform a linear transformation from the generalized Halphen variables $\big(t_1^{(m)}, t_2^{(m)}, t_3^{(m)}\big)$ to the variables $\big(\mathsf{x}^{(m)},\mathsf{y}^{(m)}, E_2^{(m)}\big)$ using \eqref{eq:xydef} and \eqref{eq:E2xy}. The generalized Halphen system then takes the form
\begin{gather}
E_2^{{(m)}^\prime} = \frac{m-2}{4m} \big( \big(E_2^{(m)}\big)^2 - \mathsf{x}^{(m)} \mathsf{y}^{(m)} \big), \nonumber\\
\mathsf{x}^{{(m)}^\prime} =\left(\frac{m-2}{2m}\right) E_2^{(m)} \mathsf{x}^{(m)} + \frac{\big(\mathsf{x}^{(m)}\big)^2}{m} -\frac{\mathsf{x}^{(m)} \mathsf{y}^{(m)}}{2}, \nonumber\\
\mathsf{y}^{{(m)}^\prime} =\left(\frac{m-2}{2m}\right) E_2^{(m)} \mathsf{y}^{(m)} - \left(\frac{m-1}{m}\right)\mathsf{x}^{(m)} \mathsf{y}^{(m)} +\frac{\big(\mathsf{y}^{(m)}\big)^2}{2} .\label{eq:xydbydtau}
\end{gather}
The first equation in \eqref{eq:xydbydtau} proves the first of the Ramanujan identities. For the higher weight forms, we simply differentiate \eqref{eq:E2kxy} and use \eqref{eq:xydbydtau} to find
\begin{gather*}
E_{2k}^{{(m)}^\prime} = (k-1)\big(\mathsf{x}^{(m)}\big)^{k-2}\mathsf{y}^{(m)} \left[ \left(\frac{m-2}{2}\right) E_2^{(m)} \mathsf{x}^{(m)} +\frac{\big(\mathsf{x}^{(m)}\big)^2}{m} - \frac{\mathsf{x}^{(m)} \mathsf{y}^{(m)}}{2}\right] \cr
\hphantom{E_{2k}^{{(m)}^\prime} =}{} + \big(\mathsf{x}^{(m)}\big)^{k-1}\left[\left(\frac{m-2}{2}\right) E_2^{(m)} \mathsf{y}^{(m)} +\left(\frac{m-1}{m}\right)\mathsf{x}^{(m)} \mathsf{y}^{(m)} + \frac{\big(\mathsf{y}^{(m)}\big)^2}{2}\right] \\
\hphantom{E_{2k}^{{(m)}^\prime}}{}= \frac{k}{2}\left(\frac{m-2}{m}\right) E_{2k}^{(m)} E_2^{(m)} - \left(\frac{k-2}{2}\right) E_4^{(m)} E_{2k-2}^{(m)} - \left(\frac{m-k}{m}\right) E_{2k+2}^{(m)} .
\end{gather*}
The second term can be resolved in multiple equivalent ways, courtesy of \eqref{eq:notFreelyGenerated}. We have thus derived the analogues of the Ramanujan identities for Hecke groups, which for fixed $ m $ and $ k \leq m $ take the stated form for any $ k'$ such that $ 2 \leq k' \leq k-1 $.
\end{proof}

It is straightforward to verify that the case $m=3$ reproduces the well known Ramanujan identities corresponding to the modular group. The above identities for $ m = 4 $ and $ 6 $ have appeared in \cite{Zudilin2003}. For Hecke groups ${\rm H}(m)$, we get analogous relations; systems of this form will be collectively referred to as Ramanujan identities.

In earlier work \cite{Raman:2018owg}, similar Ramanujan identities were derived for cusp forms associated to the Hecke group ${\rm H}(m)$. In Appendix~\ref{earlierwork} we translate between the two sets of automorphic forms~-- the Eisenstein series $ E_{2k}^{(m)} $ and the cusp forms $f_{2k}^{(m)}$~-- and relate~\eqref{eq:RamanujanIdGeneral} to the identities obtained in~\cite{Raman:2018owg}.

\section[Hyperelliptic curves, anomaly equations, and boundary conditions]{Hyperelliptic curves, anomaly equations,\\ and boundary conditions} \label{sec:CurvesAnomalies}

In this section we propose to give a geometric interpretation to the Ramanujan identities for the arithmetic Hecke groups following earlier work on the subject by \cite{Dubrovin:1994, Katz:1976, Movasati2012a, Movasati2012b}.

\subsection{Curves and anomalies}

We begin by assigning to each of the Hecke groups, an algebraic curve defined by the equation
\begin{gather*}\label{eq:DefineCurve}
y^2 = p_m(x),
\end{gather*}
where $ p_m(x) $ is a polynomial of degree $ m $ defined as
\begin{gather}\label{eq:DefineP}
p_m(x) = x^m + \sum_{k=1}^{m} x^{m-k} A_k^{(m)} .
\end{gather}

\begin{Definition}The coefficients $ A_k^{(m)} $ are chosen recursively via an {\it anomaly equation}
\begin{gather}\label{eq:ModularAnomalyEquation}
\frac{\partial A_{k}^{(m)}}{\partial E_{2}^{(m)}}=\frac{c (m-k+1)}{m} A_{k-1}^{(m)},
\end{gather}
supplemented by boundary conditions which demand that, as $ \tau \rightarrow \ii\infty $,
\begin{gather}\label{eq:CuspBC}
A_{k}^{(m)}=O\big(q^{\lfloor k / 2\rfloor}\big).
\end{gather}
We will set $ c = 1/4 $ for definiteness in \eqref{eq:ModularAnomalyEquation}, and conventionally set $ A_0^{(m)}= 1 $. The boundary condition~\eqref{eq:CuspBC} will be referred to as a cuspidal boundary condition.
\end{Definition}

We now supply a derivation of the anomaly equation. First, assign to the coefficient $ A_k^{(m)} $ a~weight $ 2k $ under the action of the relevant Hecke group; for consistency, we need to associate to~$ x $ the weight~$ 2 $, and to~$ y $ the weight~$ m $. With this assignment of weights, $ A_1^{(m)} $ has weight $ 2 $, and can thus be chosen to be proportional to $ E_2^{(m)} $
\begin{gather}\label{eq:A1vsE2}
A_1^{(m)} = c E_2^{(m)} .
\end{gather}
The $ A_k^{(m)} $ are so far characterized solely by their weight. In general, this makes them quasi-automorphic objects. Drawing from motivations relating to the sort of algebraic curves that appear in supersymmetric gauge theories \cite{Ashok:2016ewb,DHoker:1999yni,Donagi:1995cf}, we note that it is often desirable to have algebraic curves whose coefficients are purely automorphic forms. This requirement is imposed by demanding that $p_m(x)$ with shifted argument is such that the coefficients are purely automorphic
\begin{gather}\label{eq:ShiftedCurve}
\widetilde{p}_m(x):=p_m\left(x-\frac{c}{m}E_2^{(m)}\right) = x^m + \sum_{k=2}^{m} x^{m-k}\widetilde{A}_k^{(m)},
\end{gather}
where under the action of $\gamma =\left(\begin{smallmatrix}
a & b\\
c & d
\end{smallmatrix}\right) \in {\rm H}(m) $, the $ \widetilde{A}_k^{(m)} $ transform as
\begin{gather*}
\widetilde{A}_k^{(m)} (\gamma \cdot \tau)  = (c \tau + d)^{2k} \widetilde{A}_k^{(m)}(\tau), \qquad \text{with}\quad \gamma\cdot\tau = \frac{a \tau +b}{c \tau + d},
\end{gather*}
i.e., they are automorphic forms of weight $ 2k $. This would in turn imply that the coeffi\-cients~$ \widetilde{A}_k^{(m)} $ for $ k > 2 $ do not depend on the quasi-automorphic Eisenstein series $ E_2^{(m)} $. An interesting outcome of this requirement can be derived by rewriting~$p_m(x)$ as
\begin{gather*}
p_m(x) = \left(x+\frac{c}{m}E_2^{(m)}\right)^m + \sum_{k=2}^{m} \left(x+\frac{c}{m}E_2^{(m)}\right)^{m-k}\widetilde{A}_k^{(m)},
\end{gather*}
which in turn allows us to translate our requirement into a constraint of the form
\begin{gather*}
\left[\frac{\p}{\p E_2^{(m)}} - \frac{c}{m}\frac{\p}{\p x} \right]p_m(x) = 0 ,
\end{gather*}
where we now write the polynomial $ p_m(x) $ as in \eqref{eq:DefineP}. This constraint along with the relation~\eqref{eq:A1vsE2} yields
\begin{gather*}
\sum_{k=1}^{m} \left[ \frac{\p A_k^{(m)}}{\p E_2^{(m)}} -\frac{c}{m} (m-k+1) A_{k-1}^{(m)} \right] x^{m-k} = 0.
\end{gather*}
This provides a set of $ (m-1) $ equations, each constraining the dependence of the $A_k^{(m)}$ on the quasi-automorphic Eisenstein series $ E_2^{(m)} $, and each of the form
\begin{gather}\label{eq:AnomalyEquationGeneral}
\frac{\p A_k^{(m)}}{\p E_2^{(m)}} = \frac{c (m-k+1)}{m} A_{k-1}^{(m)}.
\end{gather}
Note in particular that the above argument is worked out in complete generality: it is true for all Hecke groups ${\rm H}(m)$, and applies to the coefficients of their associated hyperelliptic curves. For $ m=3 $, these kinds of equations have appeared in the context of topological string theory and superconformal gauge theories \cite{Ashok:2015cba,Ashok:2016oyh,Billo:2015pjb,Billo:2015jyt,Huang:2013eja} and are referred to as modular anomaly equations. Since the above equation constrains the dependence of the $ A_k^{(m)} $ on the quasi-automorphic Eisenstein series $ E_2^{(m)} $, we will refer to them (more generally) as anomaly equations.

The constant $ c $ is a matter of convention, but we will treat it as being universal (i.e., the same for all Hecke groups) for convenience; for the case $ m=3 $, we fix this constant to be $ c = 1/4 $ by appealing to a well-known factorisation of the elliptic curve associated to the modular group, which we discuss in Section \ref{examples}.

Integrating the anomaly equations fixes the dependence of $ A_k^{(m)} $ on the quasi-automorphic form $ E_2^{(m)} $. In order to determine $ A_k^{(m)} $ completely, we must supply a boundary condition that fixes the purely automorphic pieces. We now highlight one possible choice of boundary condition that will be useful in later sections.

Starting with the $ A_k^{(m)} $ of the lowest weight, we solve the anomaly equation, thereby ``integrating in'' any $E_2^{(m)}$ dependence. For example, on using \eqref{eq:A1vsE2}, the anomaly equation for $ A_2^{(m)} $ reads
\begin{gather*}
\frac{\p A_2^{(m)}}{\p E_2^{(m)}} = \frac{c^2 (m-1)}{m} E_2^{(m)},
\end{gather*}
which is solved by
\begin{gather*}
A_2^{(m)} = \frac{c^2 (m-1)}{2m} \big(E_2^{(m)}\big)^2 + (\text{automorphic form}).
\end{gather*}
The constants of integration are automorphic forms under the group ${\rm H}(m)$. In order to determine them, we write down every possible automorphic form consistent with considerations of weight, accompanied by coefficients that are to be determined. For $ A_2^{(m)} $, this means
\begin{gather*}
A_2^{(m)} = \frac{c^2 (m-1)}{2m} \big(E_2^{(m)}\big)^2 + a E_4^{(m)},
\end{gather*}
as for all ${\rm H}(m)$, the dimension of the space of weight $4$ forms is unity. For general $ A_k^{(m)} $, the number of terms we can write down (i.e., the number of undetermined coefficients) will by definition be as many terms as the dimension of the space $\mathfrak{m}_{2k}$, which from~\eqref{eq:Notes_Collated:1} is $ \lfloor k/2 \rfloor$.

We propose to fix these coefficients by demanding that near the cusp at $ \ii\infty $, the $A_k^{(m)}$ has a~Fourier expansion that starts at $q^{\dim \mathfrak{m}_{2k}}$. That is, as $ \tau \rightarrow \ii\infty $
\begin{gather*}\label{eq:CuspBCDefn}
A_k^{(m)} = O\big(q^{\lfloor k/2 \rfloor}\big) .
\end{gather*}
This boundary condition provides as many equations as the number of coefficients to be determined, and is consequently an unambiguous prescription. Further, by construction these $ A_k^{(m)} $ satisfy the anomaly equation. Additionally, we will see in Section~\ref{sec:Discriminants} that the choice of cuspidal boundary conditions allow us to relate the `curve' discriminant $ \Delta^{(m)} $ constructed out of the (hyper-)elliptic curve to the `automorphic' discriminant $ \Delta_m $ defined solely with reference to the ring of quasi-automorphic forms for ${\rm H}(m)$ in an interesting manner.

\subsection{Examples}\label{examples}
\subsubsection[${\rm H}(3)$]{$\boldsymbol{{\rm H}(3)}$}

In this section we show explicitly that the solutions to the modular anomaly equations along with the cuspidal boundary conditions are completely consistent with the usual parametrisation of the elliptic curves. The elliptic curve associated to the modular group is
\begin{gather}\label{eq:H3Curve}
y^2 = x^3 + A_1^{(3)} x^2 + A_2^{(3)} x + A_3^{(3)}.
\end{gather}
Notice that the assignment of weights here implies that the coefficients $ A_2^{(3)} $ and $ A_3^{(3)} $ have weights $ 4 $ and $ 6 $ respectively. On solving the anomaly equations and using the cuspidal boundary conditions, we find
\begin{gather*}
A_1^{(3)} = c E_2^{(3)}, \qquad
A_2^{(3)} = \frac{c^2}{3} \big( \big(E_2^{(3)}\big)^2 - E_4^{(3)} \big), \\
 A_3^{(3)} = \frac{c^3}{27} \big( \big(E_2^{(3)}\big)^3 - 3 E_2^{(3)} E_4^{(3)} + 2 E_6^{(3)} \big).
\end{gather*}

{\bf Factorization of elliptic curves.} The polynomial defining the elliptic curve admits a~well-known factorisation
\begin{gather}\label{eq:H3CurveFactorized}
y^2 = \prod_{k=1}^{3}  (x+s_k ),
\end{gather}
where, following \cite{Harnad:1999hi}, we define the $s_k$ to be logarithmic derivatives of Jacobi $ \theta $-constants\footnote{The Jacobi $\theta$-function is defined as follows
\begin{gather*}
\theta\left[ \begin{smallmatrix} a \\ b \end{smallmatrix} \right](v | \tau)=\sum_{n \in \mathbb{Z}} q^{\left(n-\frac{a}{2}\right)^{2}} \mathrm{e}^{2 \pi \mathrm{i}\left(n-\frac{a}{2}\right)\left(v-\frac{b}{2}\right)} .
\end{gather*}
The $\theta$-constants are defined by setting $v=0$ and $\theta_{2} \equiv \theta\left[ \begin{smallmatrix} 1 \\ 0 \end{smallmatrix} \right], \theta_{3} \equiv \theta\left[ \begin{smallmatrix} 0 \\ 0 \end{smallmatrix} \right], \theta_{4} \equiv \theta\left[ \begin{smallmatrix} 0 \\ 1 \end{smallmatrix} \right]$.}
\begin{gather*}
s_1 = \frac{1}{\ii\pi} \frac{\dd{}}{\dd{\tau}} \log \theta_2(\tau), \qquad
s_2 = \frac{1}{\ii\pi} \frac{\dd{}}{\dd{\tau}} \log \theta_3(\tau), \qquad
s_3 = \frac{1}{\ii\pi} \frac{\dd{}}{\dd{\tau}} \log \theta_4(\tau).
\end{gather*}
The $ A_k^{(3)} $ are symmetric polynomials in the $ s_k $, i.e.,
\begin{gather*}
A_1^{(3)}  = s_1 + s_2 + s_3, \qquad
A_2^{(3)}  = s_1 s_2 + s_2 s_3 + s_3 s_1, \qquad
A_3^{(3)}  = s_1 s_2 s_3,
\end{gather*}
and the requirement of consistent assignment of weights leads us to conclude that the roots $ s_k $ have weight $ 2 $. Let us focus on the first of the anomaly equations, which yields $A_1^{(3)} = c E_2^{(3)}$ in terms of a sum of three $s_k$. As we know, a $ \tau $-derivative raises the weight of a modular (function or form) by two units. Using this explicit factorized solution, we can solve for $A_1^{(3)}$ as
\begin{gather}\label{A1eqn1}
A_1^{(3)} = \frac{1}{\ii\pi} \frac{\dd{}}{\dd{\tau}} \log \theta_2(\tau) \theta_3(\tau) \theta_4(\tau), \\
\label{A1eqn2} \hphantom{A_1^{(3)}}{} = \frac{1}{\ii\pi} \frac{\dd{}}{\dd{\tau}} \log \eta(\tau)^3 = \frac{1}{4}E_2^{(3)} .
\end{gather}
In going from equation \eqref{A1eqn1} to equation \eqref{A1eqn2} we have made use of the following identity that relates the product of the three $\theta$-constants to the Dedekind $\eta$-function\footnote{The Dedekind $\eta$-function is defined as the following product
$\eta(\tau) = q^{\frac{1}{24}}\prod\limits_{n=1}^{\infty} (1-q^n)$.}
\begin{gather*}
\theta_2(\tau)\theta_3(\tau)\theta_4(\tau) = 2\eta^3(\tau).
\end{gather*}
In the last equality in equation \eqref{A1eqn2} we have used the fact that the modular discriminant (which equals the curve discriminant in this case) is related to the $\eta$-function by the relation
\begin{gather*}
\Delta = (2\pi)^{12}\eta^{24}(\tau) ,
\end{gather*}
and that the Eisenstein series is related to the modular discriminant by the relation
\begin{gather*}
E_2^{(3)}(\tau) = \frac{1}{2\pi \ii} \frac{\dd}{\dd\tau} \log\Delta .
\end{gather*}
This fixes the constant $ c = \frac{1}{4} $, and motivates the simplifying assumption that $ c = \frac{1}{4} $ for the anomaly equations associated to all ${\rm H}(m)$. The other $A_k^{(3)}$ can similarly be checked to be consistent with those that arise from the factorized curve.

\subsubsection[${\rm H}(4)$]{$\boldsymbol{{\rm H}(4)}$}

The case of ${\rm H}(4)$ works out in much the same way. We begin with the quartic curve
\begin{gather*}
y^2 = x^4 + \sum_{k=1}^{4} x^{4-k} A_k^{(4)} .
\end{gather*}
The $A_k$ satisfy our anomaly equation \eqref{eq:AnomalyEquationGeneral} with $ m = 4 $
\begin{gather*}
\frac{\partial A_k^{(4)} }{\partial E_2^{(4)} } = \frac{5-k}{16} A_{k-1}^{(4)}.
\end{gather*}
Together with the cuspidal boundary conditions, we find the following solutions to the anomaly equations
\begin{gather*}
A_1^{(4)} = \frac{1}{4} E_2^{(4)}, \qquad
A_2^{(4)}  = \frac{3}{128} \big( \big(E_2^{(4)} \big)^2 - E_4^{(4)} \big), \\
A_3^{(4)} = \frac{1}{1024} \big( \big(E_2^{(4)} \big)^3 - 3 E_2^{(4)} E_4^{(4)} + 2 E_6^{(4)} \big), \\
A_4^{(4)}  = \frac{1}{65536} \big( \big(E_2^{(4)} \big)^4-6 \big(E_2^{(4)} \big)^2 E_4^{(4)} +8 E_2^{(4)} E_6^{(4)} +\big(E_4^{(4)} \big)^2 -4 E_8^{(4)} \big).
\end{gather*}
As in the case of H$(3)$, it is possible to present a factorized form for the quartic curve
\begin{gather}\label{eq:H4CurveFactorized}
x^4 + \sum_{k=1}^{4} A_k^{(4)} x^{4-k} = \prod_{k = 1}^{4} (x+s_k) ,
\end{gather}
from which it follows that $A_k^{(4)}$ are elementary symmetric polynomials
\begin{gather}
A_1^{(4)} = s_1 + s_2 + s_3 + s_4 , \nonumber\\
A_2^{(4)} = s_1 s_2+s_3 s_2+s_4 s_2+s_1 s_3+s_1 s_4+s_3 s_4, \nonumber\\
A_3^{(4)} = s_1 s_2 s_3+s_1 s_4 s_3+s_2 s_4 s_3+s_1 s_2 s_4, \nonumber\\
A_4^{(4)} = s_1 s_2 s_3 s_4.\label{AvsSk}
\end{gather}
The $s_k$ that factorize the quartic polynomial in \eqref{eq:H4CurveFactorized} are given as follows
\begin{alignat}{3}
& s_1 = \frac{1}{2\pi\ii} \frac{\dd{}}{\dd{\tau}} \log \theta_2(2\tau), \qquad&&
s_2 = \frac{1}{2\pi\ii} \frac{\dd{}}{\dd{\tau}} \log \theta_3(2\tau), &\nonumber\\
& s_3 = \frac{1}{2\pi\ii} \frac{\dd{}}{\dd{\tau}} \log \theta_3(\tau), \qquad&&
s_4 = \frac{1}{2\pi\ii} \frac{\dd{}}{\dd{\tau}} \log \theta_4(\tau).&\label{skintermsoftheta}
\end{alignat}
This is verified by explicitly comparing the Fourier expansions on both sides of~\eqref{AvsSk}.\footnote{The expressions for the $s_k$ in \eqref{skintermsoftheta} were independently derived by~\cite{Govindarajan2012}.}

\subsection{General solution to the anomaly equation}

To recapitulate, we started by associating to each Hecke group a curve that was elliptic for $m\leq 4$ and hyperelliptic for $m > 4$. We then derived an anomaly equation by insisting that in a shifted form of the curve, any dependence of the coefficients $A_k^{(m)} $ on the quasi-automorphic form $E_2^{(m)}$ disappeared. We proposed a natural choice of boundary conditions that allowed us to unambiguously determine the purely automorphic pieces, i.e., the constants of integration that are obtained after ``integrating in'' the $E_2^{(m)}$ dependence. Apart from an overall constant that varies with the height $ m $ of the Hecke group, the~$ A_k^{(m)} $ are found to be proportional to the same combinations of the Eisenstein series. The first few $A_k^{(m)}$ are given below
\begin{gather*}
A_1^{(m)} \propto E_2^{(m)}, \\
A_2^{(m)} \propto \big[ \big(E_2^{(m)}\big)^2 - E_4^{(m)} \big], \\
A_3^{(m)} \propto \big[ \big(E_2^{(m)}\big)^3 - 3 E_2^{(m)} E_4^{(m)} + 2 E_6^{(m)} \big], \\
A_4^{(m)} \propto \left[ \big(E_2^{(m)}\big)^4 - 6 \big(E_2^{(m)}\big)^2 E_4^{(m)} + 8 E_2^{(m)} E_6^{(m)} + \big(E_4^{(m)}\big)^2 - 4 E_8^{(m)} \right], \\
A_5^{(m)} \propto \big[ \big(E_2^{(m)}\big)^5 - 10 \big(E_2^{(m)}\big)^3 E_4^{(m)} + 20 \big(E_2^{(m)}\big)^2 E_6^{(m)}   \\
 \hphantom{A_5^{(m)} \propto}{} + 5 E_2^{(m)} \big(\big(E_4^{(m)}\big)^2 - 4 E_8^{(m)}\big) - 4 E_4^{(m)} E_6^{(m)} + 8 E_{10}^{(m)} \big],
\end{gather*}
and so on. One can show that
\begin{gather*}
A_k^{(m)} = {m-1\choose k-1}\frac{1}{k 4^k m^{k-1}} \big[ \big(E_2^{(m)}\big)^k + \cdots \big],
\end{gather*}
where the constant of proportionality is a simple consequence of the anomaly equation.

\subsection{Discriminants}\label{sec:Discriminants}
We have defined a set of (hyper-)elliptic curves with coefficients determined by an anomaly equation together with cuspidal boundary conditions. Since these curves $y^2 = p_m(x)$ are specified by the polynomials $ p_m(x) $, it is natural to consider the discriminant of the polynomial, defined in the usual way as the resultant of the polynomial~$ p_m(x)$ and its derivative~$ p_m'(x) $. It is a~polynomial in the $A_k^{(m)}$ with integer coefficients, which we will denote by $\Delta^{(m)}$ and refer to as the curve discriminant.

Since the coefficients $ A_k^{(m)} $ transform as quasimodular forms of weight~$ (2k) $, we then ask what the weight of the curve discriminant might be. Let us suppose that the polynomial admits a~factorisation
\begin{gather*}
p_m(x) = \prod_{k=1}^{m} (x + s_k),
\end{gather*}
in which case the $ A_k^{(m)} $ are easily seen to be symmetric polynomials in the $ s_k $. For the weights to be consistently defined, the $ s_k $ must transform as weight $ 2 $ quasimodular forms. In terms of the factorised representation of the polynomial, the discriminant is defined as
\begin{gather*}
\Delta^{(m)} = \prod_{i \neq j} (s_i - s_j),
\end{gather*}
which makes it clear that the weight of the algebraic discriminant is $w = 2m(m-1) $.

Now, we must also use the information that the ring of quasi-modular forms for Hecke groups~${\rm H}(m)$ with $ m > 3 $ is not freely generated, and that there are non-trivial relations between the Eisenstein series of higher weight. It is easily verified that for $E_{2k}^{(m)} $ (with $ k > 3 $) we have
\begin{gather*}
 E_{2k}^{(m)}\big(E_4^{(m)}\big)^{k-3} = \big( E_6^{(m)} \big)^{k-2},
\end{gather*}
which we can plug into the definition of the curve discriminant in terms of resultants. This motivates the following conjecture:
\begin{Conjecture}
The curve discriminant for the $($hyper-$)$elliptic curves associated to the Hecke group ${\rm H}(m)$ is given by
\begin{gather*}
\Delta^{(m)} = a_m \frac{\big(\big(E_4^{(m)}\big)^3-\big(E_6^{(m)}\big)^2\big)^{(m-1)(m-2)/2}}{\big(E_4^{(m)}\big)^{(m-1)(m-3)}}, \qquad\text{with}\quad a_m = \frac{2^{2-m(m+1)}}{ m^{m(m-2)}}.
\end{gather*}
\end{Conjecture}

Let us try and massage the above expression into something more familiar. We rewrite the automorphic forms in terms of the standard hauptmodul $J_{(m)}$ and its derivatives using the formulae in Appendix~\ref{earlierwork} (see equation~\eqref{E2kintermsofJ})
\begin{gather}\label{eq:DeltaJf}
\Delta^{(m)}
=a_m \left[\frac{\big(J_{(m)}'\big)^m}{J_{(m)}^{m-1} \big(J_{(m)}-1\big)^{m/2}}\right]^{(m-1)} = \begin{cases}
a_m \big(f_{2m}^{(m)}\big)^{(m-1)} &\text{for} \ m \in 2\mathbb{Z}, \\
a_m \big(f_{4m}^{(m)}\big)^{(m-1)/2} &\text{for} \ m \in 2\mathbb{Z}+1.
\end{cases}
\end{gather}

Finally, the automorphic discriminant (which we will encounter later as well) is defined in \cite[Theorem~2(ii)]{Doran:2013npa} as $ \Delta_m := f^{(m)}_{2L}$ with $ L = \operatorname{lcm}(2,m) $, where $ f^{(m)}_{2k} $ are cusp forms defined in~\eqref{eq:f2kDefinition}. It is natural to wonder what the relationship between $\Delta^{(m)} $, the curve discriminant of the (hyper-)elliptic curve which we have determined earlier to have weight $w = 2m(m-1)$, and the automorphic discriminant $\Delta_m$ is. From \cite[Theorem~2(ii)]{Doran:2013npa} we see that the weight of~$ \Delta_m $ is
\begin{gather*}
w_m = \begin{cases}
2m &\text{for} \ m \in 2\mathbb{Z}, \\
4m &\text{for} \ m \in 2\mathbb{Z} + 1.
\end{cases}
\end{gather*}
From \eqref{eq:DeltaJf} we find that the algebraic and automorphic discriminants are related as
\begin{gather*}
\left(\frac{\Delta^{(m)} }{a_m}\right)^{w_m} =  ( \Delta_m  )^{w},
\end{gather*}
which serves as a strong, non-trivial consistency check on the web of relationships we have uncovered. In particular, it confirms the correctness the (hyper-)elliptic curve and the anomaly equation, and justifies the use of cuspidal boundary conditions.

\subsection{Gauss--Manin connections}

Here we discuss a geometric interpretation of the Ramanujan identities following \cite{Dubrovin:1994, Katz:1976, Movasati2012a,Movasati2012b}. The goal here will be to associate to the Ramanujan identities~-- being as they are a set of ordinary differential equations~-- a vector field on the parameter space of the elliptic curve in~\eqref{eq:H3CurveFactorized}. We then do the same for H$ (4) $. For this, we use the notion of a Gauss--Manin connection, which formalizes the following observation: the variation of an elliptic integral~-- defined using some basis of differentials~-- with respect to a parameter~$t$ can be written as a linear combination of period integrals
\begin{gather*}
\dd  \left( \oint \Pi_a \right) = \sum_{b} \mathcal{A}_{ab} \left( \oint \Pi_b \right).
\end{gather*}
The coefficients $ \mathcal{A} $ form a $ 2 \times 2 $ matrix, and for multiple such parameters $ \{ t_k \}_{k=1}^{m} $ we define the differential form
\begin{gather*}
\mathcal{A} = \sum_{k = 1}^{m} \mathcal{A}^{(k)} \dd t_k,
\end{gather*}
so the variation of the period integrals with respect to these parameters is captured in the equations
\begin{gather*}
\nabla \Pi = \sum_{k=1}^{m} \big( \mathcal{A}^{(k)} \Pi \big) \dd t_k =: \mathcal{A} \cdot \Pi.
\end{gather*}
More properly, $ \mathcal{A} $ should be viewed as a differential $ 1 $-form on the parameter space of the elliptic curve. The matrix $ \mathcal{A} $ is referred to as a Gauss--Manin connection, and in \cite[Proposition 6]{Movasati2012b} this Gauss--Manin connection was computed explicitly for the $m=3$ case.

We now study what happens to this basis of differential $ 1 $-forms as we vary $ \tau $, the modular parameter of the underlying torus. This leads us to define the Ramanujan vector field
\begin{gather*}
{\rm R} := \frac{1}{2\pi\ii} \frac{\dd}{\dd{\tau}}.
\end{gather*}
In terms of variations of the parameters that appear in the curve (which are the Eisenstein series), we have
\begin{gather}\label{eq:RamanujanTau}
{\rm R} = \sum_{k=1}^{m} \left(\frac{1}{2\pi\ii} \frac{\dd}{\dd{\tau}} E_{2k}^{(m)} \right) \frac{\p}{\p E_{2k}^{(m)}} .
\end{gather}
The portion in parentheses above may be replaced in accordance with the Ramanujan identi\-ties~\eqref{eq:RamanujanIdGeneral}.

We can contract this differential $ 1 $-form on the parameter space of the elliptic curve with a vector from the same space. The Ramanujan identities define the vector field \eqref{eq:RamanujanTau} on the parameter space of the elliptic curve, and we define its contraction with the connection as
\begin{gather*}
\nabla_{\rm R} = \iota_{{\rm R}} \nabla,
\end{gather*}
using the rule
\begin{gather*}
\frac{\p}{\p t_k} ( \dd t_\ell ) = \delta_{k\ell}.
\end{gather*}

We now ask how $ \nabla_{{\rm R}} $ acts on a basis of differential $ 1 $-forms associated to our elliptic curve, and in \cite[Proposition~7]{Movasati2012b} it was demonstrated that
\begin{gather*}
\nabla_{\rm R} \Pi = \mathcal{A}_{{\rm R}} \Pi,
\end{gather*}
with $ \mathcal{A}_{\rm R}$ determined by explicit calculation. Let us quickly review the manner in which this computation was performed, albeit in a shifted form of the curve. The coefficient of the quadratic term in the elliptic curve can be set to zero through a shift of the form
\begin{gather*}
x \rightarrow x - \frac{1}{3} A_1^{(3)}.
\end{gather*}
Then, the curve \eqref{eq:H3Curve} in terms of the Eisenstein series takes its Weierstrass normal form
\begin{gather*}
y^2 = \widetilde{p}_3(x) = x^3 - \frac{E_4^{(3)}}{48} x + \frac{E_6^{(3)}}{864}.
\end{gather*}
The coefficients of the polynomial $ \widetilde{p}_3 $ are precisely the automorphic forms $ \widetilde{A}_k^{(3)} $ that we encountered in \eqref{eq:ShiftedCurve}. Our goal will be to determine the action of the Ramanujan vector fields on the period integrals; this computation is built out of constituents that have the general form
\begin{gather}\label{eq:VariationPiGeneral}
\frac{\p}{\p E_{2k}^{(3)}} \oint \frac{x^\ell \dd{x}}{y} = \oint \frac{\dd{x}}{\widetilde{p}_3y} \left( -\frac{x^\ell}{2} \frac{\p \widetilde{p}_3}{\p E_{2k}^{(3)}} \right) .
\end{gather}
At this stage, a happy consequence of using the shifted polynomial $ \widetilde{p}_3 $ is that we can conclude by construction
\begin{gather*}
\frac{\p}{\p E_{2}^{(3)}} \oint \frac{x^\ell \dd{x}}{y} = 0,
\end{gather*}
i.e., the period integrals do not vary in the direction $ \p_{E_2^{(3)}} $ as the curve $y^2 = \widetilde{p}_3(x)$ carries no dependence on $ E_2^{(3)} $. For the other components of the Ramanujan vector field $ \p_{E_{2k}^{(3)}} $ (for $ k \in \lbrace 2,3 \rbrace $) and each independent differential form $ \frac{x^{\ell} \dd{x}}{y} $ (for $ \ell \in \lbrace 0,1 \rbrace $), we follow a technique of~\cite{Movasati2012b} and look for polynomials $ \alpha $ and $ \beta $ that satisfy the constraint
\begin{gather*}
-\frac{x^\ell}{2} \frac{\p}{\p E_{2k}^{(3)}} \widetilde{p}_3 = \alpha \frac{\dd{\widetilde{p}_3 }}{\dd{x}} + \beta \widetilde{p}_3.
\end{gather*}
Once determined, we plug this into the variation in question~\eqref{eq:VariationPiGeneral} and after some elementary manipulations, it is then easy to see that the result of the variation with respect to $ E_{2k}^{(3)} $ is simply
\begin{gather*}
\frac{\p}{\p E_{2k}^{(3)}} \oint \frac{x^\ell \dd{x}}{y} = \oint \frac{\dd{x}}{y} \left( 2\frac{\dd{\alpha}}{\dd{x}} + \beta \right).
\end{gather*}
Earlier work has used this technique to establish the following theorem:
\begin{Theorem}[Movasati \cite{Movasati2012b}]
The Gauss--Manin connection corresponding to the Ramanujan vector field associated to the modular group rotates the canonical basis of differential $ 1 $-forms on the elliptic curve as
\begin{gather*}
\mathcal{A}_{{\rm R}} = \begin{pmatrix}
-\dfrac{E_2^{(3)}}{12} & 1 \vspace{1mm}\\
-\dfrac{E_4^{(3)}}{144} & \dfrac{E_2^{(3)}}{12}
\end{pmatrix}.
\end{gather*}
\end{Theorem}
\begin{Remark}This computation may also be performed in the original (unshifted) basis, and in that case the result is
\begin{gather*}
\mathcal{A}_{{\rm R}} = \begin{pmatrix}
0 & 1 \\
0 & 0
\end{pmatrix},
\end{gather*}
which is precisely the result of \cite{Movasati2012b} up to a sign.
\end{Remark}

\subsubsection{Height four}
While the holomorphic differential is still $ \frac{\dd{x}}{y} $~-- and this is true for all hyperelliptic curves~-- the differential $ \frac{x \dd{x}}{y} $ is no longer a good candidate as it has a simple pole at infinity, making it a~differential of the third kind~\cite{Broedel:2017kkb}. Instead, a valid differential of the second kind is given by
\begin{gather*}
\Pi_2 = \frac{x^2 \dd{x}}{y}.
\end{gather*}
This choice of differential does not have a pole at infinity. We use the following convenient basis of differential $ 1 $-forms for an elliptic curve defined by a quartic polynomial
\begin{gather*}
\Pi = \left(\begin{matrix}
\dfrac{\dd{x}}{y} \vspace{1mm}\\ \dfrac{x^2 \dd{x}}{y}
\end{matrix}\right)  .
\end{gather*}

\begin{Theorem}The Gauss--Manin connection corresponding to the Ramanujan vector field associated to H$ (4) $ rotates the above basis of differential $ 1 $-forms on the elliptic curve as
\begin{gather*}
\mathcal{A}_{{\rm R}} = \begin{pmatrix}
-\dfrac{E_2^{(4)}}{4} & 0 \vspace{1mm}\\
-\dfrac{E_6^{(4)}}{512} & \dfrac{E_2^{(4)}}{4}
\end{pmatrix}.
\end{gather*}
\end{Theorem}

\begin{proof}The proof proceeds by explicit computation. The Ramanujan vector field is given by
\begin{gather*}
{\rm R} = \frac{1}{2\pi\ii} \frac{\dd}{\dd{\tau}}  = \sum_{k=1}^{4} \left(\frac{1}{2\pi\ii} \frac{\dd}{\dd{\tau}} E_{2k}^{(4)} \right) \frac{\p}{\p E_{2k}^{(4)}},
\end{gather*}
and as before, the expression in parentheses may be replaced in accordance with the Ramanujan identities, giving
\begin{gather*}
{\rm R}  = \frac{1}{8} \big(\big(E_2^{(4)}\big)^2 - E_4^{(4)}\big) \frac{\p}{\p E_2^{(4)}} + \frac{1}{2} \big(E_2^{(4)} E_4^{(4)} - E_6^{(4)}\big) \frac{\p}{\p E_4^{(4)}} \\
\hphantom{{\rm R}  =}{} + \frac{1}{4} \big(3 E_2^{(4)} E_6^{(4)} - 2 \big(E_4^{(4)}\big)^2 - E_8^{(4)} \big) \frac{\p}{\p E_6^{(4)}} + \big(E_2^{(4)} E_8^{(4)} - E_4^{(4)} E_6^{(4)}\big) \frac{\p}{\p E_8^{(4)}}.
\end{gather*}
The basis of differential $ 1 $-forms will rotate into itself under the action of the Ramanujan vector field. After a shift of the form
\begin{gather*}
x \rightarrow x - \frac{1}{4} A_1^{(4)}
\end{gather*}
the curve (in terms of Eisenstein series) takes the form
\begin{gather*}
y^2 = \widetilde{p}_4(x) = x^4 - \frac{3 E_4^{(4)}}{128} x^2 + \frac{E_6^{(4)}}{512} x + \frac{\big(E_4^{(4)}\big)^2 - 4E_8^{(4)}}{65536}.
\end{gather*}
Employing the technique of solving for polynomials $ \alpha $ and $ \beta $ allows us to determine the effect of the Ramanujan vector field on the basis of differential $ 1 $-forms. We find that
\begin{gather*}
\nabla_{\rm R} \Pi_1  = -\frac{E_2^{(4)}}{4} \Pi_1, \qquad
\nabla_{\rm R} \Pi_2  = -\frac{E_6^{(4)}}{512} \Pi_1 + \frac{E_2^{(4)}}{4} \Pi_2.
\end{gather*}
Alternatively, we may write down the following connection equation
\begin{gather*}
\nabla_{{\rm R}} \Pi = \mathcal{A}_{{\rm R}} \Pi ,
\end{gather*}
where $ \mathcal{A}_{{\rm R}} $ is as claimed.
\end{proof}

\section{Chazy equations}\label{sec:Chazy}

In this section we derive the Chazy equation and its higher-order
analogues, each of them canonically associated to a set of Ramanujan
identities that are in a one-to-one correspondence with the Hecke
groups.

We will also show that like the Chazy equation, its higher-order generalizations also possess the
Painlev\'e property. In particular, the Chazy equation and its
generalizations possess negative resonances, which in turn naively imply
the instability of linear perturbations about its solutions. We will show that the negative resonances vanish ``on-shell,'' i.e., when the Chazy equation is satisfied. This demonstrates the stability of these solutions against linear
perturbations.

Let us quickly review the relation between the Chazy equation and the Ramanujan identities corresponding to the modular group. The Ramanujan identities for H$ (3) $ take the following form
\begin{gather*}
E_2^{{(3)}^\prime}  = \frac{1}{12} \big( \big(E_2^{(3)}\big)^2 - E_4^{(3)}\big) , \\
E_4^{{(3)}^\prime}  = \frac{1}{3} \big( E_2^{(3)} E_4^{(3)} - E_6^{(3)}\big) , \\
E_6^{{(3)}^\prime}  = \frac{1}{2} \big( E_2^{(3)} E_6^{(3)} - \big(E_4^{(3)}\big)^2\big) .
\end{gather*}
A well-known strategy (outlined for example in \cite{Chanda:2015baj}) consists of using the above equations to find a differential equation satisfied by the weight $2$ Eisenstein series $E_2^{(3)}$, which we will denote by~$ y $. This is done straightforwardly by elimination and we find the following equation:
\begin{gather}\label{eq:Chazy3}
2 y^{(3)} -2 y y'' +3 y'^2 = 0,
\end{gather}
where  $y^{(m)}$ is $m$-th derivative of $y$ with respect to $\tau$ (see, e.g., equation~\eqref{eq:AHSII_Revised_V3:1}). Up to rescalings, this nonlinear third-order differential equation is known as the Chazy equation \cite{CLARKSON1996225}. It originally arose in the study of third-order ordinary differential equations having the Painlev\'e property. We shall return to a detailed study of this property in Section \ref{sec:Painleve}. Before this, we derive higher-order Chazy equations that are derived straightforwardly from the Ramanujan identities.

\subsection{Higher-order Chazy equations}
Now that we have a procedure for deriving the Chazy equation corresponding to Ramanujan's identities, it is natural to ask: if the Ramanujan identities admit a generalization to the case of Hecke groups, what do the corresponding Chazy equations $ C_m $ look like?

{\bf An example: height four.}
Let us consider the Ramanujan identities \eqref{eq:RamanujanIdGeneral} for $m=4$. Using the symbol $y$ to once again denote $E_2^{(4)}$ in analogy with the case of the modular group, and following the procedure outlined in the previous section yields a new, fourth-order analogue of the Chazy equation\footnote{This equation was independently derived in \cite{Govindarajan2012}.}
\begin{gather*}
C_4 \colon \ 4 y^{(4)} -10 y^{(3)} y +6 y^2 y'' -9 y (y')^2+12 y' y'' = 0.
\end{gather*}
The structure of the Ramanujan identities is uniform across all heights, so it is natural to expect that the order of the differential equation matches the number of generalized Ramanujan identities, which is the same as the height $m$ of the Hecke group ${\rm H}(m)$ in question.

{\bf Chazy equations for $\boldsymbol{{\rm H}(m)}$.}
By following the same logic one can construct higher-order analogues of the Chazy equation for ${\rm H}(m)$. Below, we list the next two members of this family, for future reference
\begin{gather}
\nonumber C_5 \colon \ 300 y^{(5)} -1350 y^{(4)} y +1932 y^{(3)} y^2-882 y^3 y'' +920 (y'')^2 \\
\hphantom{C_5 \colon\ }{} +1323 y^2 (y')^2-168 (y')^3+y' \big(620 y^{(3)} -2772 y y'' \big) = 0,\nonumber \\
\nonumber C_6 \colon \   9 y^{(6)} -63 y^{(5)} y +158 y^{(4)} y^2-168 y^{(3)} y^3-156 y (y'')^2 \\
\nonumber  \hphantom{C_6 \colon \ }{} +48 y (y')^3+\big(78 y^{(3)} +64 y^4\big) y'' -(y')^2 \big(48 y'' +96 y^3\big) \\
\hphantom{C_6 \colon \ }{}  -y' \big(3 y^{(4)} +68 y^{(3)} y -240 y^2 y'' \big) = 0.\label{eq:Chazy6}
\end{gather}
Thus, we find constructively that:
\begin{Proposition}Each Hecke group ${\rm H}(m)$ is associated to an order~$ m $ nonlinear ordinary differential equation. Further, each term in the Chazy equation corresponding to ${\rm H}(m)$ has weight $ 2m+2 $.
\end{Proposition}

\begin{proof}Consider the Ramanujan identities corresponding to ${\rm H}(m)$, a set of $ m $ first-order differential equations in $ m $ variables. The above proposition follows by differential elimination.
\end{proof}

\subsection[${\rm H}(m)$ orbits]{$\boldsymbol{{\rm H}(m)}$ orbits}\label{sec:HeckeOrbits}
In the previous section, we have demonstrated constructively that $ y=E_2^{(m)} $ satisfies a higher-order Chazy equation, which for Hecke group ${\rm H}(m)$ is a nonlinear ordinary differential equation of order~$ m $. This allows us to generalise \cite[equation~(8)]{Takhtajan:1992qb} in the following way:
\begin{Proposition} The automorphic discriminant $ \Delta_m $ is the Hirota $ \tau $-function for the Chazy equation $ C_m $.
\end{Proposition}
This follows rather trivially following the representation of $ y=E_2^{(m)} $ in terms of the `automorphic' discriminant $ \Delta_m $ as given in \cite[Theorem~2(ii)]{Doran:2013npa}
\begin{gather*}
y = \frac{1}{2\pi\ii} \frac{\dd{}}{\dd{\tau}} \log \Delta_m .
\end{gather*}

We now characterise general solutions to the Chazy equation~$ C_m $. This is inspired by \cite[Lemma~3]{Takhtajan:1992qb} and proceeds by rewriting~$ C_m $ as a linear combination of automorphic forms of a~given weight. Let us see how this works in the case of the usual Chazy equation corresponding to~${\rm H}(3) $, thereby reviewing \cite[Theorem~2]{Takhtajan:1992qb}. The expression
\begin{gather*}
Z = E_2^{{(3)}^\prime} - \frac{1}{12}\big(E_2^{(3)}\big)^2
\end{gather*}
is a weight $ 4 $ modular form. We want to generate forms of higher weight, and a well-known procedure to do this is to use the Ramanujan--Serre derivatives~\cite{Zagier2008} that send
\begin{gather*}
{\rm D}\colon \ \mathfrak{m}_{k} \rightarrow \mathfrak{m}_{k+2}.
\end{gather*}
Explicitly, this derivative takes the form (with our normalizations)
\begin{gather*}
{\rm D} = \frac{1}{2\pi\ii} \frac{\dd{}}{\dd{\tau}} - \frac{k}{12} E_2^{(3)},
\end{gather*}
and one can check that with this definition $ {\rm D}E_4^{(3)} = -\frac{1}{3} E_6^{(3)} $ and $ {\rm D}E_6^{(3)} = - \frac{1}{2} \big(E_4^{(3)}\big)^2 $. We now act with $ {\rm D} $ on $ Z $ until we get a weight $ 8 $ form, since each term in the Chazy equation we are interested in has weight $ 8 $. There are essentially two terms we can write down, and we consider a linear combination of them
\begin{gather*}
a {\rm D}^2Z + b Z^2,
\end{gather*}
which is guaranteed to be an automorphic form of weight $ 8 $. Finally, we can check that the Chazy equation can be written as
\begin{gather*}
C_3\colon \ {\rm D}^2 Z + 2 Z^2 =0.
\end{gather*}
We \looseness=-1 have thus demonstrated that $ C_3 $ can be written as a linear combination of weight $ 8 $ automorphic forms; thus, under the action of a $ \gamma=\left(\begin{smallmatrix}
a & b\\
c & d
\end{smallmatrix}\right) \in {\rm H}(3) $, each term $ x $ in $ C_3 $ will transform as
\begin{gather*}
x \mapsto (c\tau+d)^8 x.
\end{gather*}
With this, we can conclude that all H$ (3) \cong {\rm SL}(2,\mathbb{Z}) $-orbits of $ y $ solve the Chazy equation.

Further, the space of weight $8$ forms is $1$-dimensional while the above procedure generates two weight $8$ forms. It follows that there must be a relation between them, and so some linear combination of the two must vanish. This combination is precisely the Chazy equation. We now generalise these statements to all Hecke groups ${\rm H}(m)$.

\begin{Theorem}All ${\rm H}(m)$ orbits of $ E_2^{(m)} $ are solutions to $ C_m $. That is, if $ y=E_2^{(m)} $ is a solution to $ C_m $, then so is $ \widetilde{y} = E_2^{(m)}(\gamma\cdot\tau) $, where $ \gamma \in {\rm H}(m)$.
\end{Theorem}

\begin{proof}The strategy of our proof is similar: for ${\rm H}(m)$ we will define the weight $ 4 $ form
\begin{gather*}
Z = E_2^{{(m)}^\prime} - \left( \frac{m-2}{4m} \right) \big(E_2^{(m)}\big)^2,
\end{gather*}
and invoke a result of \cite{Raman:2018owg} that defines an appropriate analogue of the Ramanujan--Serre derivatives for the Hecke group~${\rm H}(m)$. In our present normalization this takes the form
\begin{gather}\label{eq:DGeneral}
{\rm D} = \frac{1}{2\pi\ii} \frac{\dd{}}{\dd{\tau}} - \frac{k}{2} \left(\frac{m-2}{2m}\right) E_2^{(m)}.
\end{gather}
It will turn out that under the action of a $ \gamma=\left(\begin{smallmatrix}
a & b\\
c & d
\end{smallmatrix}\right) \in {\rm H}(m) $, each term $x$ that appears in $ C_m $ will transform as
\begin{gather*}
x \mapsto (c\tau+d)^{2m+2} x.
\end{gather*}
This can be explicitly verified for Hecke groups with small values of $m$. For example, we find
\begin{gather*}
C_4 \colon \   {\rm D}^3 Z + 6 Z {\rm D}Z = 0 , \\
C_5 \colon \  3 {\rm D}^4 Z + 26 Z {\rm D}^2 Z + 20 ({\rm D}Z )^2 + 12 Z^3 = 0 , \\
C_6 \colon \  {\rm D}^5 Z +12 Z {\rm D}^3 Z + 24 {\rm D}^2Z {\rm D} Z + 32 Z^2 {\rm D}Z = 0,
\end{gather*}
and so on. In each of these cases, the proof goes through as before.
\end{proof}

Similarly, it is easy to understand why it is reasonable to expect that $ C_m $ is zero.
\begin{Remark}It is known that for the Hecke group ${\rm H}(m)$ there are $ m $ of generators of the ring of quasi-automorphic forms $ {\mathbf {\mathcal R}_m} $ \cite{Doran:2013npa}. Thus, at weight $ (2m+2) $, all forms that span this vector space will be products of forms of lower weight and their modular covariant derivatives. Any weight $ (2m+2) $ form must be expressible as a linear combination of these products/derivatives. We may thus conclude on general grounds that these Chazy equations are statements of linear dependence.
\end{Remark}

\subsection{Relation to Maier's equation}\label{sectionMaier}

An important point to keep in mind is that the quasi-automorphic forms do not generate a free polynomial algebra for $m>3$. This is immediately obvious from the definition of the $E_{2k}^{(m)}$ in~\eqref{eq:E2kxy}; for instance it is true for all $m>3$ that
\begin{gather}\label{E4E8}
E_{4}^{(m)} E_{8}^{(m)} = \big(E_6^{(m)}\big)^2 .
\end{gather}
Similar relations hold for the higher weight forms as well. Given the way we derived the generalized Chazy equation, this immediately suggests the existence of a lower-order differential equation satisfied by $y=E_2^{(m)}$ for all $m$ by using such identities; in this section we show that this is indeed the case and in fact the differential equation we obtain is identical to the one obtained in~\cite{Maier}.

Let us see this in detail. Consider the action of the Ramanujan--Serre derivative \eqref{eq:DGeneral} on $E_4$. The following identity is
straightforwardly verified
\begin{gather*}
(m-2) E_4^{(m)} {\rm D}^2 E_4^{(m)} = (m-3) \big({\rm D}E_4^{(m)}\big)^2+\frac{(m-2)^2}{2m} \big(E_4^{(m)}\big)^3 .
\end{gather*}
Upon defining $ Y = \frac{(m-2)^2} {16m^2}E_4^{(m)} $, we obtain the following relation of Maier~\cite{Maier}
\begin{gather*}
(m-2) Y \mathrm{D}^{2} Y-(m-3)(\mathrm{D} Y)^{2}-8 m Y^{3} = 0.
\end{gather*}
This shows the complete equivalence of the Ramanujan identities for ${\rm H}(m)$ and Maier's equation. In particular, this implies the following theorem.
\begin{Theorem}For all ${\rm H}(m)$, the weight $ 2 $ quasi-automorphic form $ y = E_2^{(m)} $ satisfies the following third-order nonlinear differential equation
\begin{gather}
y{}{'''} \left(2y^2-\frac{8 my'}{(m-2)}\right)+y'' \left(\frac{24 y y'}{(m-2)}-2y^3\right) \nonumber\\
\qquad{} +\frac{8 m(m-3) (y'')^2}{(m-2)^2}+3 y^2(y')^2-\frac{4 (m+6)(y')^3}{(m-2)} = 0 .\label{MaierChazy}
\end{gather}
\end{Theorem}

\begin{proof} With the identity \eqref{E4E8}, the set of Ramanujan identities \eqref{eq:RamanujanIdGeneral} for $ k = \lbrace 1, 2, 3 \rbrace $ becomes a closed system of first-order ordinary differential equations. By elimination, we recover the above differential equation for $ y=E_2^{(m)} $.
\end{proof}

This form of the Maier equation will prove to be very useful when we perform a Painlev\'e analysis on the solutions to the differential equation.

\section{Painlev\'e analysis of the Chazy equations}\label{sec:Painleve}

We now test the Chazy equations for the Painlev\'e property, which roughly corresponds to a~statement that the only movable singularities of these differential equations are poles. We begin with a brief primer on stability analysis and go on to apply the methods of~\cite{Conte:1989pi,Conte:1993pp} to the higher-order Chazy equations as well as the Maier equation.

\subsection{A primer on stability}

We begin with a brief review of the stability analysis of nonlinear differential equations. Consider a nonlinear ordinary differential equation of order~$n$. The analysis due to Painlev\'e~\cite{Painleve1902} involves expanding the independent variable near the singular point~-- this ``nearness'' is pa\-ra\-met\-rized by a small parameter~$\alpha$~-- and expanding the dependent variable as a formal power series in $\alpha$. The method due to Kowalevskaya \cite{Sophie1890} on the other hand involves a Laurent expansion of the dependent variable around the singular point, leading to algebraic equations for the coefficients.

These tests were subsequently refined in \cite{Ablowitz:1980ca}, where the methods of Kowalevskaya and Painlev\'e were combined. It was demonstrated that the Painlev\'e property is a necessary condition for the differential equation to be integrable. This method proposes that the local solution around the singularity is of the Frobenius form
\begin{gather} \label{eq:1}
 y(\tau) = \sum_{i=0}^\infty y_i (\tau-\tau_0)^{i-a} .
\end{gather}
In \cite{Ablowitz:1980ca} a set of criteria were identified, and differential equations that satisfied all of them were said to possess the Painlev\'e property. The first of these criteria is that $a$ is a positive integer, which in effect reduces the solution to the Laurent series form. This part of the test amounts to studying the indicial equation, and to go further, one linearizes the equation around the movable singularity. Let the original equation be
\begin{gather*}
K\big[y\big] = 0,
\end{gather*}
where $K[y]$ is a polynomial function of $y$, $y'$, etc.~up to the $n^{{\rm th}}$ derivative, then the linearized equation is obtained as
\begin{gather}\label{lineardefn}
\frac{\dd{}}{\dd{\epsilon}} K[y+\epsilon w]\bigg\vert_{\epsilon=0} = 0.
\end{gather}
Substituting the Frobenius form \eqref{eq:1} into the above equation, one can equivalently write the linearized equation as
\begin{gather*}
K_{{\rm L}}\left[\frac{\dd{}}{\dd{\tau}} \right] w(\tau) = 0 ,
\end{gather*}
for some polynomial $K_{\rm L}$ whose coefficients are given by the $y_i$. We assume the following ansatz for the linearized solution:
\begin{gather} \label{eq:2}
 w(\tau) = w_0 (\tau-\tau_0)^{-a} + \sum_{i=1}^\infty w_i (\tau-\tau_0)^{i-a} .
\end{gather}
This ansatz supposes that the linearized equation also has a Frobenius series solution. For a~linear differential equation of order~$n$, the Frobenius analysis tells us that the series is a~solution if the coefficient of $(\tau-\tau_0)^{i-a-n}$ vanishes for each~$i$. This condition gives a recursion relation, which determines $w_i$ in terms of $w_j$ with $j<i$, and the parameters appearing in the linear differential equation. We write this equation as
\begin{gather*}
 P(i) w_i - f (\tau_0, a, y_k; w_1,\dots, w_{i-a} ) = 0.
\end{gather*}
Let $r_1\leq r_2\leq\cdots \leq r_n$ be roots of the polynomial $P(i)$; then we can determine $w_i$ in terms of $w_j$ (with $j<i$) as long as $i \neq r_1$. When $i=r_1$ then we have $P(i)=0$ and we end up with a condition $f(\tau_0, a, y_k, w_1, \dots, w_{i-a})=0$. If this condition is satisfied then $w_{r_1}$ is indeterminate and the linearized equation is said to pass the Painlev\'e test. If, however, $f(\tau_0, a, y_k, w_1, \dots, w_{i-a})\not=0$ then we have a logarithmic branch; in this case we say that the linearized equation does not possess the Painlev\'e property.

If the linearized equation passes the Painlev\'e test at $i=r_1$, we continue the test for $i=r_k$ for $k=\{2,\dots , n\}$. The linearized equation is said to possesses the Painlev\'e property if and only if it passes the Painlev\'e test at each root of $P(i)$. These criteria are collectively known as the Ablowitz--Ramani--Segur (ARS) stability conditions~\cite{Ablowitz:1980ca}. If the linearized equation satisfies the ARS stability conditions then we say that the original nonlinear equation possesses the Painlev\'e property.

The procedure we have outlined above assumes that the linearized equation has positive resonances, i.e., the solutions to the linearized equation have singularities that are less severe than those of the nonlinear equation. This is not true in general; in fact, the Chazy equation is a well-known counter-example. While the Chazy equation passes the first condition of having only movable poles, the linearized equation turns out to have negative resonances, and in these situations the techniques of \cite{Ablowitz:1980ca} are insufficient. Luckily, the analysis of negative resonances has been carried out in \cite{Conte:1989pi,Conte:1993pp} and has the added advantage of being applicable to nonlinear partial differential equations, thereby subsuming the analysis of \cite{Weiss:1983pt}.

We now turn to the notion of stability for systems with negative resonances. In a nutshell, the argument of \cite{Conte:1989pi,Conte:1993pp} is that if the coefficients of the negative resonances vanish identically when the zeroth-order nonlinear equation is satisfied, we are permitted to conclude that the equation possesses the Painlev\'e property.

\subsection{The Painlev\'e property}

We will now consider the Painlev\'e analysis of the Chazy equations $C_m$, examples of which are presented in equations~\eqref{eq:Chazy3} to~\eqref{eq:Chazy6}, as well as the Maier equation in \eqref{MaierChazy}. These equations are nonlinear ordinary differential equations and all these equations satisfy the first criterion of ARS with $a=1$, implying that every nonlinear equation has a solution with simple movable poles. Before discussing the general case we begin by reviewing the analysis of the original Chazy equation~$C_3$ in~\eqref{eq:Chazy3}, following~\cite{Fordy:1991nd}.

In order to illustrate the procedure, we first seek a solution to the equation in the Frobenius form, i.e., \eqref{eq:1}. The indicial equation gives $a=1$ for the ``maximal'' case: the case in which all the terms in the Chazy equation scale in the same fashion as we scale $\tau\to \lambda \tau$ and $y \to \lambda^a y$. The integrality of $a$ ensures that the Chazy equation passes the first criterion of ARS. We then proceed to determine the coefficients $y_i$ by recursively solving the equation \eqref{eq:Chazy3}. Some low-order coefficients are
\begin{gather*}
 y_0=-12 ,\qquad y_1=y_2=y_3=0 .
\end{gather*}
To see if this solution is stable against perturbation we substitute $y(\tau) = y_s(\tau) + \epsilon w(\tau)$ in the Chazy equation, where $y_s(\tau)$ is the solution to the Chazy equation. The linearized equation is obtained by picking up terms linear in~$\epsilon$, as in \eqref{lineardefn}, and we find
\begin{gather}\label{eq:10}
 3y'(\tau) w'(\tau) - y''(\tau) w(\tau) - y(\tau) w''(\tau) + w'''(\tau) = 0 .
\end{gather}
Substituting the ansatz \eqref{eq:2} into this equation, one finds that the linearized solutions of equation~\eqref{eq:10} has poles of order higher than those of the solution to the Chazy equation, in particular poles of order two and three in addition to the usual simple pole. As discussed earlier, this vio\-la\-tes the criterion of \cite{Ablowitz:1980ca}, which implicitly assumes that the resonances are less singular than the original solution. In order to perform the stability analysis we now outline a strategy to circumvent this difficulty \cite{Fordy:1991nd}. This involves a~Frobenius expansion of the solution written in terms of a function which reflects the fact that the solution has movable poles. This is done in two steps.

First we define a function $\phi(\tau)$, which parametrizes the singular manifold when $\phi(\tau)=0$. The solution, however, is written in terms of a Frobenius series in another function $\chi(\tau)$, related to $\phi(\tau)$ as follows
\begin{gather*}
\chi(\tau)= \frac{2\phi \phi'}{2(\phi')^2 - \phi\phi''} .
\end{gather*}
This provides us with a germ which is formally independent of the function defining the singular manifold. The function $\chi(\tau)$ satisfies a Riccati type equation
\begin{gather*}
 \chi'(\tau) = 1 - \frac{1}{2} S \chi^2(\tau),
\end{gather*}
where the Schwarzian $S$ is defined to be
\begin{gather*}
S = \frac{\phi'''(\tau)}{\phi(\tau)} -\frac{3}{2} \left(\frac{\phi''(\tau)}{\phi'(\tau)}\right)^2 .
\end{gather*}
We now make the following ansatz for the leading order solution
\begin{gather}\label{leadingorderansatz}
y(\tau) = \sum_{i=0}^{\infty} y_i \left(\chi(\tau)\right)^{i-a} ,
\end{gather}
and find that $a=1$ as before but now find the following solutions for the coefficients
\begin{gather}\label{yiC3leading}
y_0 = -12 , \qquad y_1=0 , \qquad y_2= 2S , \qquad y_3 =0 ,
\end{gather}
and so on. Note that we now get a non-zero value for $y_2$, unlike the earlier (naive) Laurent expansion. For the next-to-leading order we now use the following ansatz
\begin{gather} \label{eq:12}
 w(\tau) = \sum_{i=0}^{\infty}w_i (\chi(\tau))^{i-b} .
\end{gather}
Substituting this in the linearized Chazy equation \eqref{eq:10} gives $b=\{3, 2, 1\}$~-- which shows the existence of higher-order poles for the solution to the linearized equation compared to the solution of the Chazy equation itself. However, when we substitute the coefficients $y_i$ in \eqref{yiC3leading}, we find that the coefficients of
the resonances at $b=3$ and $b=2$ vanish identically. This implies in turn that these resonances do not destabilise the solution of the Chazy equation.

We now carry out a similar analysis on the general third-order differential equation that is equivalent to the Maier equation and that we obtained in \eqref{MaierChazy}. Substituting the ansatz \eqref{leadingorderansatz} and solving for the exponent, we find $a=1$. For the leading coefficients, we now find
\begin{gather}\label{Maiercoeff}
y_0 = -\frac{4m}{m-2} , \qquad y_1 = 0 , \qquad y_2 = \frac{2m}{3(m-2)} S , \qquad y_3 = 0 .
\end{gather}
At the next to leading order analysis we find $b=2$, which shows the presence of higher-order poles for the linearized equations compared to the solution of the Maier equation itself. But exactly as for the Chazy equation, we find that when we substitute the coefficients $y_i$ in \eqref{Maiercoeff}, the coefficients of the higher resonance vanish identically and implies that the resonances do not destabilize the solution of the Maier equation.

We thus conclude that the Maier equation possesses the Painlev\'e property. Further, from the discussion in Section~\ref{sectionMaier}, it is clear that the higher-order Chazy equations $C_m$ are entirely equivalent to the Maier equation when the algebraic dependence of the forms $ \big\lbrace E_{2k}^{(m)} \big\rbrace_{k=2}^{m} $ (see equation \eqref{eq:notFreelyGenerated}) is taken into account. We therefore conclude that each $C_m$ also possesses the Painlev\'e property. We have explicitly verified this for $3\leq m\leq 10$ and the analysis of \cite{Conte:1989pi,Conte:1993pp} leads us to the following proposition.
\begin{Proposition}The Maier equation in \eqref{MaierChazy} as well as the higher-order Chazy equation $ C_m $ possess the Painlev\'e property. The leading order ansatz $y(\tau)$ defined in \eqref{leadingorderansatz} satisfies both sets of differential equations for the following values of the coefficients
\begin{gather} \label{eq:16}
 y(\tau) = \frac{2m}{m-2}\left(-\frac{2}{\chi(\tau)} + \frac{S}{3} \chi(\tau) + \frac{2S^2}{45} \chi(\tau)^3 + \frac{11S^3}{945}\chi(\tau)^5+\cdots \right) .
\end{gather}
\end{Proposition}
The analysis of the next-to-leading order ansatz \eqref{eq:12} shows that the Chazy equation has negative resonances, with $b \in \{1, 2,\dots,m\}$ for $C_m$. The number of resonances is equal to the order of the linearized equation and therefore the set of resonances is maximal. Exactly as we saw in the case of $C_3$ as well as the Maier equation, we find that when evaluated on the leading order solution, the coefficients of these resonances vanish identically, guaranteeing that all these equations have the Painlev\'e property. It is interesting to observe that while the Maier equation fixes the first two coefficients in the expansion in \eqref{eq:16}, it is the Painleve analysis of the higher-order Chazy equations that allows one to fix the sub-leading terms in the ansatz.

\section{Discussion}

We have studied several interesting properties of automorphic forms and associated nonlinear differential equations associated to the Hecke group ${\rm H}(m)$. We now discuss some future directions for research motivated by these developments, with a special emphasis on applications relevant to gauge theories, integrability, and string theories.

Hecke groups have appeared as strong-weak duality groups acting on coupling constants in certain supersymmetric gauge theories~\cite{Ashok:2015cba}. These coupling constants appear in what is called the low energy effective action of the gauge theory. Given this, we expect the effective action and other calculable quantities to be expressible in terms of the automorphic forms discussed in this work. While this has been done for arithmetic Hecke groups in~\cite{Ashok:2015cba,Ashok:2016oyh} it would be interesting to carry it out for the other Hecke groups. From the gauge theory perspective what it amounts to is a resummation of the observables that allows one to probe the theory at all values of the coupling, which is a desirable feature in any quantum field theory.

The differential equations $C_m$ we have uncovered merit further investigation from the point of view of integrability, since we have demonstrated that each $ C_m $ possesses the Painlev\'e property, a necessary condition for a differential equation to be integrable.

The congruence subgroups of ${\rm SL}(2,\mathbb{Z})$ also arise in the context of state counting in four dimensional $N=4$ supersymmetric string theories \cite{Dabholkar:2005dt}. Of particular interest is the generating function that counts states which preserve half of the $N=4$ supersymmetry; this generating function is given in terms of a denominator formula which contains either $\eta$-products or $\eta$-quotients.\footnote{In fact, there exists a larger class of $N=4$ superstring models whose duality group contains the Fricke involution $\tau\to -\frac{1}{N\tau}$, in addition to the subgroup of ${\rm SL}(2,\mathbb{Z})$~\cite{Persson:2015jka}.} It is not difficult to check that the solution to the Chazy equation for $m=3,4$ can be written in terms of logarithmic derivatives of the very same $\eta$-products that appear in these generating functions.\footnote{Since $E_2^{(m)}$, which solves the Chazy equation, is proportional to the function $A_1^{(m)}$ that appears in the curve associated to the Hecke group, this can be seen from the factorized form of the curve in \eqref{eq:H3CurveFactorized} and \eqref{eq:H4CurveFactorized}, and some well known $\theta$-function identities.} It would be interesting to understand this relation further, and in particular understand how the Hecke groups organise the degeneracies of the supersymmetric states in these string theories.

\appendix

\section{Generalized Halphen system and Fourier expansions}\label{app:Fourier}

In this section we collect a few results about the Halphen system and the automorphic forms $E_{2k}^{(m)}$ associated to ${\rm H}(m)$. The generalized Halphen system was introduced in \eqref{eq:GenHalphen}. In this section, we provide explicit solutions to this system of differential equations following \cite[Theorem 3]{Doran:2013npa}. In terms of the parameters $ (a,b,c) $ introduced in \eqref{eq:abcHalphen}, we have the solution
\begin{gather}
t_1^{(m)}(\tau) =\frac{1}{\alpha_m} (a-1) z Q(a, b; z) {}_2F_1(1-a, b;1 ;z ) {}_2F_1(2-a, b;2 ; z),\nonumber \\
t_2^{(m)}(\tau) - t_1(\tau) = \frac{1}{\alpha_m} Q(a, b; z) {}_2F_1(1-a, b;1 ;z )^2,\nonumber \\
t_3^{(m)}(\tau) -t_1(\tau) = \frac{1}{\alpha_m} z Q(a, b; z) {}_2F_1(1-a, b;1 ;z )^2.\label{eq:HalphenGenSol}
\end{gather}
The function ${}_2F_1$ is the Gauss hypergeometric function while the function $Q(a ,b; z)$ is given by
\begin{gather*}
Q(a, b; z) = \frac{\ii \pi (1-b)}{2\sin (\pi b) \sin (\pi a)} (1-z)^{b-a} .
\end{gather*}
The parameter $z$ is related to the standard hauptmodul $J_{(m)}$ of the Hecke group ${\rm H}(m)$ by the formula
\begin{gather*}
z = \frac{1}{1-J_{(m)}}.
\end{gather*}
Here we have rescaled the solutions in \cite{Doran:2013npa} by an overall constant $\alpha_m$, which we will fix momentarily. For the Hecke group ${\rm H}(m)$ we now work out the Fourier expansions of the solutions of the generalized Halphen equations $t_k^{(m)}$ near the cusp at ${\rm i}\infty$. These, in turn, determine the $q$-expansions for the Eisenstein series $E_{2k}^{(m)}$ of the Hecke group via \eqref{eq:E2kxy}.

An important ingredient in the Fourier expansion of the solutions to the generalized Halphen system is the Fourier expansion of the standard hauptmodul corresponding to the Hecke group in question. This is obtained by solving a Schwarzian differential equation order-by-order about the point $ \tau = \ii\infty $ following the prescriptions in \cite{Doran:2013npa, Raman:2018owg}. This yields
\begin{gather*}
J_{(m)} = \frac{1}{dq}+\frac{4+3m^2}{8m^2} +\frac{dq}{1024m^2}\big(69m^4-8m^2-48\big)\\
\hphantom{J_{(m)} =}{} +\frac{d^2 \left(27 m^6-116 m^4+16 m^2+64\right) q^2}{3456 m^6}+ \cdots,
\end{gather*}
where $d$ is defined following \cite{Doran:2013npa}. First define the integers $a'$, $b'$, $c'$, $d'$ such that
\begin{gather*}
\frac{a'}{b'} = \frac{3m-2}{4m}, \qquad \frac{c'}{d'} = \frac{3m+2}{4m} .
\end{gather*}
Then $d$ is defined to be the following product
\begin{gather*}
d = b' d' \prod_{k=1}^{b'-1}
\left(2-2\cos\frac{2\pi k}{b'} \right)^{-\frac{1}{2}\cos\frac{2\pi a'}{b'}}\prod_{\ell=1}^{d'-1} \
\left(2-2\cos\frac{2\pi \ell}{d'} \right)^{-\frac{1}{2}\cos\frac{2\pi c'}{d'}} .
\end{gather*}
For the arithmetic cases of $m=3, 4,$ and~$6$, $d$ takes integer values and is equal to $1728$, $256$, and $108$ respectively. We now demand that the expansion for $t_2^{(m)}$ begins with unit coefficient; this uniquely fixes the coefficient $\alpha_m$ to be
\begin{gather*}
\alpha_m = -\frac{(3m+2)\sec\frac{\pi}{m}}{8m} .
\end{gather*}
With our conventions now made fully explicit, it is easy to check that~\eqref{eq:HalphenGenSol} solves the generalized Halphen system. We can now use~\eqref{eq:E2kxy} and~\eqref{eq:E2xy} to construct Fourier expansions of the Eisenstein series we use in this paper. For example, the first few Eisenstein series have the following Fourier expansions
\begin{gather*}
E_2^{(m)} = 1-\frac{(m-2)^2}{8 m^2}(qd)-\frac{(m (m+12)-44) (m-2)^2}{512 m^4}(qd)^2 + \cdots, \\
E_4^{(m)} = 1 +\frac{\big(m^2-4\big) }{4 m^2}(qd)+ \frac{\big(7 m^4-88 m^2+240\big)}{256 m^4} (qd)^2+ \cdots, \\
E_6^{(m)} = 1 -\frac{\big(m^2+12\big) }{8 m^2}(qd) + \frac{\big({-}31 m^4+152 m^2+912\big)}{512 m^4}(qd)^2+\cdots, \\
E_8^{(m)} = 1-\frac{\big(m^2+4\big) }{2 m^2}(qd) -\frac{\big(m^4-168 m^2-368\big)}{128 m^4} (qd)^2 + \cdots.
\end{gather*}
These expansions prove useful in implementing the cuspidal boundary conditions and finding the solutions to the anomaly equations satisfied by the $A_k^{(m)}$.

\section{Ramanujan identities for cusp forms}\label{earlierwork}

Recall the definition \cite[Theorem 2(i)]{Doran:2013npa} of the cusp form $f_{2k}^{(m)}$ of a Hecke group ${\rm H}(m)$
\begin{gather}\label{eq:f2kDefinition}
f_{2k}^{(m)} = (-1)^k \big(J_{(m)}'\big)^k J_{(m)}^{1 -k}\big(J_{(m)}-1\big)^{\lceil\frac{k}{2}\rceil-k}.
\end{gather}
Here $J_{(m)}$ is the standard hauptmodul of the Hecke group ${\rm H}(m)$, that in turn solves a Schwarzian differential equation. In order to relate these to the Eisenstein series $E_{2k}^{(m)}$ in~\eqref{eq:E2kxy}, we recall from \cite[Theorem 4(iii)]{Doran:2013npa} that the solutions to the generalized Halphen system can be written down in terms of the standard hauptmodul of the corresponding Hecke group
\begin{gather*}
\mathsf{x} = \frac{J_{(m)}'}{J_{(m)}}\qquad \text{and} \qquad \mathsf{y} = \frac{J_{(m)}'}{J_{(m)}-1} .
\end{gather*}
Substituting these into the definition of $E_{2k}^{(m)}$, we obtain
\begin{gather}\label{E2kintermsofJ}
E_{2k}^{(m)} = \big(J_{(m)}'\big)^k J_{(m)}^{1 -k}(J_{(m)}-1)^{-1}.
\end{gather}
This yields the following simple relation between the Eisenstein series and the cusp form
\begin{gather}\label{eq:Evsf}
(-1)^k E_{2k}^{(m)} = f_{2k}^{(m)} \big(J_{(m)}-1\big)^{d_{2k}},
\end{gather}
where $ d_{2k} $ is defined as \cite[Theorem~2(i)]{Doran:2013npa}
\begin{gather*}
d_{2k} = k - \bigg\lceil\frac{k}{2}\bigg\rceil -\bigg\lceil\frac{k}{m}\bigg\rceil .
\end{gather*}
It is related to the dimension of the space of weight $(2k)$ automorphic forms $ \mathfrak{m}_{2k} $ (for $k\le m$) as
\begin{gather}
 \label{eq:Notes_Collated:1}
\dim \mathfrak{m}_{2k} = d_{2k} +1 = k - \left\lceil \frac{k}{2} \right\rceil = \left\lfloor \frac{k}{2}\right\rfloor.
\end{gather}
In \eqref{eq:Notes_Collated:1}, we have used the fact that $k \le m$, and the floor and ceiling functions are defined as
\begin{gather*}
\lfloor x \rfloor = \max \lbrace n \in \mathbb{Z} \colon n \leq x \rbrace, \qquad
\lceil x \rceil = \min \lbrace n \in \mathbb{Z} \colon n \geq x \rbrace.
\end{gather*}
From these definitions it is clear that
\begin{gather*}
\lceil x\rceil-\lfloor x\rfloor= \begin{cases} 0 & \text{for} \ x \in \mathbb{Z}, \\ 1 & \text{for} \ x \notin \mathbb{Z}, \end{cases}
\end{gather*}
and in particular, for $ n \in \mathbb{Z} $, we have $ \lfloor n \rfloor = \lceil n \rceil = n$. On using~\eqref{eq:Evsf} the Ramanujan identities conjectured in \cite[Section~3]{Raman:2018owg} may be directly related to the ones derived in the previous subsection.

\subsection*{Acknowledgements}

We would like to thank Suresh Govindarajan for discussions and for collaboration during an early stage of the project. We also thank Renjan John for helpful comments on an earlier version of the manuscript, and the anonymous referees for valuable comments and feedback. MR acknowledges support from the Infosys Endowment for Research into the Quantum Structure of Spacetime. This research was supported in part by the International Centre for Theoretical Sciences (ICTS) during a visit for participating in the program~-- Quantum Fields, Geometry and Representation Theory (Code: ICTS/qftgrt/2018/07).

\pdfbookmark[1]{References}{ref}
\LastPageEnding

\end{document}